\documentstyle[12pt]{article} 
\input epsf

\begin{document}
\begin{titlepage}
\thispagestyle{empty}
\title{Taylor's Series and Dispersion Relation Analyses of the Vector Pion Form Factor and
 their Comparison  with  Perturbative and Non Perturbative Calculations}
\vspace{4.0cm}
\author{Tran N. Truong \\
\small \em Centre de Physique Th{\'e}orique, 
{\footnote {unit{\'e} propre 014 du
CNRS}}\\ 
\small \em Ecole Polytechnique \\
\small \em F91128 Palaiseau, France}

\date{September  1998}

\maketitle

\begin{abstract}
The first three coefficients of the Taylor's series expansion of the 
vevtor pion form factor  as a function of the momentum transfer 
 are evaluated using the
experimental data on the pion form factor and the P-wave $\pi\pi$ phase shifts. The real
part of the form factor as a function of energy is also calculated by dispersion relation.
Comparisons there results  with  Chiral Perturbation Theory  and unitarized
models are given.

\end{abstract}
\end{titlepage}

The purpose of this note is three-fold. First, a systematic procedure is given to
calculate the coefficients of the Taylor's series expansion of the form factor around the
origin $s=0$ in terms of experimental data. Second the real parts of the form factor for a
wider range of energy are calculated using also  experimental data. 
 Third,  these results are compared with those given by
existing theoretical calculations in order to evaluate their reliability.  It turns out
that only models  satisfying the elastic unitarity, giving rise to the $\rho$ resonance,
are valid. Chiral Perturbation calculations at one and two-loop disagree, at very low
energy,  with the Taylor's series expansion and,   at moderate energy, with the dispersion
relation results.

 Chiral Perturbation Theory (ChPT) \cite{holstein,Weinberg, GL1, GL2} is a
well-defined perturbative procedure
 allowing one to calculate systematically low energy phenomenon involving
soft pions. It is now widely used to analyze the low energy pion physics not only
when the interaction is not  strong but also in the presence of the
resonance as long as the energy region of interest is sufficiently far from the
resonance. In this scheme, the
unitarity relation is satisfied perturbatively order by order.

In this note we want to examine  critically the ChPT approach and wish to emphasize
that unless that this calculation was followed by a unitarisation procedure, its results for
a number of physical processes would not be meaningful. This is so because in any
perturbative scheme, it is important that the magnitude of the calculated term should be
much larger than that of the same order  which is uncalculable because of  its 
non-perturbative origin. To make this point clear, let us examine the one-loop ChPT
calculation of the vector pion form factor
$V(s)$ \cite{GL1, GL2}:

\begin{equation}
                 V^{pert.}(s) = 1 +\frac{s}{s_{R}}\ + {1\over 96\pi^2f_\pi^2}
((s-4m_\pi^2)
 H_{\pi\pi}({s}) + {2s\over 3} ) \label{eq:pertv}
\end{equation}
where $f_\pi=0.93 GeV$ and the r.m.s. radius of the vector form factor is related to $s_R$
by the definition $V ^{'}(0) =\frac{1}{6} r_V^2 = 1/s_{R}$. The function
$H_{\pi\pi}({s})$ is given by:
\begin{equation}
   H_{\pi\pi}(s) = (2 - 2 \sqrt{s-4m_\pi^2\over s}\ln{{\sqrt{s}+\sqrt{s-4m_\pi^2}\over
2m_\pi}})+i\pi\sqrt{s-4m_\pi^2\over s} \label{eq:H}
\end{equation}
for$ s>4m_\pi^2$; for other values of s, $H_{\pi\pi} (s)$ can be obtained by analytic
 continuation. 

From this definition, it is clear that the third term on the r.h.s. of Eq.
(\ref{eq:pertv}), being an analytic function with the same singularity as $V(s)$, can be
expanded as a Taylor's series at $s=0$ with a leading term behaving as $s^2$ at small $s$.
For small $s$  it is a reasonable
approximation to take only the first few terms of this series.

The unitarised version of Eq. (\ref{eq:pertv}), obtained by the inverse amplitude, the
Pad{\'e} approximant  or the N/D methods, is given by \cite{Truong1, Truong3}:
 \begin{equation}
         V(s) = \frac{1} {1 -s/s_{R} - {1\over 96\pi^2f_\pi^2}\{(s-4m_\pi^2)
 H_{\pi\pi}({s}) + {2s/3}\}} \label{eq:vu}
\end{equation}

It is obvious that Eq. (\ref{eq:vu}) has the Breit-Wigner resonance character
while that from Eq. (\ref{eq:pertv}) does not, although their amplitude and  first
derivative are identical at $s=0$. Furthermore, if the parameter $s_{R}$ was fixed by the
 the r.m.s. radius, the $\rho$ mass would come out to be slightly low compared measured
$\rho$ mass. Neglecting this last problem at the
moment, the Taylor's series expansion around $s=0$ reveals that Eq. (\ref{eq:vu}) gives
rise to a coefficient of the $s^2$  term as 
$(1/s_{R})^2\simeq 4.0 GeV^{-4}$ which is much larger than that coming from the
  third term of Eq. (\ref{eq:pertv}), $1/(960\pi^2m_\pi^2 f_\pi^2) \simeq 0.63
GeV^{-4}$. This is the signal of the failure of the perturbation method. While it is
difficult to detect the presence of this term using experimental data at low energy, 
this failure would show up at higher energy and with more accurate data.

 It could be argued, however, that this discrepancy could come from the unitarisation scheme,
because Eq. (\ref{eq:vu})  fits the data only  approximatively.  It is therefore desirable 
to have a direct  proof that this discrepancy is real and only depends on the  experimental
data.

Because the vector pion form factor $V(s)$ is an analytic function with a cut from
$4m_\pi^2$ to
$\infty$, the $n^{th}$ subtracted dispersion relation for $V(s)$ reads: 
\begin{equation}
V(s)=a_0+a_1s+...a_{n-1}s^{n-1}+ \frac{s^{n}}{\pi}\int_{4m_\pi^2}^\infty
\frac{ImV(z)dz}{z^{n}(z-s-i\epsilon)}
\label{eq:ff1} 
\end{equation}
where $n\geq 0$ and, for simplicity, the series around the origin is considered. The
polynomial on the R.H.S. of this equation will be referred in the following as the
subtraction constants and the last term on the R.H.S. as the dispersion integral (DI). The
evaluation of DI as a funtion of $s$ will be done later. 
  Notice that
$a_n=V^n(0)/n!$ is the coefficient of the Taylor series expansion for
$V(s)$, where
$V^n(0)$ is the nth derivative of
$V(s)$ evaluated at the origin. The condition for  Eq. (\ref{eq:ff1}) to be valid 
was  that, on the real positive s axis, the
limit $s^{-n}V(s)\rightarrow 0$ as $s\rightarrow \infty$. By the Phragmen Lindeloff
theorem, this limit would also be true in any direction in the complex s-plane  and hence
it is straightforward to prove Eq. (\ref{eq:ff1}). The coefficient $a_{n+m}$ of the Taylor's 
series  is given by:
\begin{equation}
a_{n+m} = \frac{1}{\pi}\int_{4m_\pi^2}^\infty
\frac{ImV(z)dz}{z^{(n+m+1)}}\label{eq:an}
\end{equation}
where $m\geq 0$. The meaning of this equation is clear: under the above stated assumption,
not only the coefficient $a_n$ can be calculated but all other coefficients $a_{n+m}$
can also be calculated.

 Below the inelastic threshold, from unitarity of the
S-matrix,
$V(s)$ must have the phase of the P-wave elastic
$\pi\pi$ phase shifts \cite{watson}. From the available experimental data, the inelastic
effect will manifest only at an energy above $1.3 GeV$ \cite{oller}. Therefore, below this
energy one has: 
\begin{equation}
ImV(z)  
                        =\mid V(z) \mid\sin\delta(z) \label{eq:eu} 
\end{equation}
where $\delta$ is the strong elastic P-wave $\pi\pi$ phase shifts. (There is an ambiguity
of $\pm$ sign on the RHS of Eq. (\ref{eq:eu}); the choice of the sign $+$ was made is due
to the normalisation $V(0)=1$ and the phenomenology, e.g. the rms pion radius is
positive). Because the real and imaginary parts are related by  dispersion relation, it
is important to know accurately $ImV(z)$ over a large  energy region. Below 1.3 GeV,
$ImV(z)$ can be determined accurately because the modulus of the vector form factor
\cite{barkov,aleph} and the corresponding P-wave $\pi\pi$ phase shifts are well measured
\cite{proto, hyams, martin}.

 Using the experimental data
on the pion form factor and the corresponding $\pi\pi$ phase shifts, 
$ImV(z)$ with experimental errors is given in Fig. 1. Similarly in Fig. 2, $ReV(s)$ with
experimental errors is given using the expression 
$ReV(s)=\mid V(s) \mid \cos\delta$. 

 One  first shows how the coefficients of
the Taylor's series around $s=0$ can be evaluated in terms of $ImV(s)$ then later, the
experimental 
$ImV(s)$ and
$ReV(s)$ will be used to test the validity of various theoretical model calculations .

 Following the usual definition, $
V(s) = 1 + \frac{1}{6} <r_V^2>s+c s^2 + d s^3+...$
one has $<r_V^2>=6a_1, c=a_2, d=a_3$ etc.
 If one makes a weak assumption that $V(s)/s \rightarrow 0$ as $s\rightarrow \infty$, using
Eq. (\ref{eq:an}) one gets the following results:
\begin{equation}
<r_V^2> = 0.44\pm 0.015 fm^2; c = 3.90\pm 0.10 GeV^{-4}; d = 9.70\pm 0.40 GeV^{-6}
\label{eq:rvn}
\end{equation}
where the upper limit of the integration is taken to be $1.7 GeV^2$. From the 2 $\pi$
threshold to 
$0.5 GeV$ the experimental data on the  the phase shifts are either poor or
unavailable, an extrapolation procedure has to be used. For this purpose the results of
models 1 and 2, to be discussed later, were used for $ImV(s)$. They contribute, rspectively,
5\%, 15\% and 30\% to the $a_1, a_2$ and $a_3$ sum rules.

  Because of the assumed high energy
behavior, one cannot calculate here the pion charge
$a_0$.
 From these results the radius of convergence for the Taylor's series is (much) less than $1
GeV^2$. In fact, a reasonable approximation for this series is the Taylor's series expansion
of
$V(s)\simeq 1/(1-a_1s)$. This result  is not surprising because it is the zero width
approximation for Eq. (\ref{eq:vu}). Away from the $\rho$ resonance, it is  a  better
approximation  than the ChPT calculations for the vector pion form factor.

If one was willing to make a stronger assumption that the form factor vanished
asymptotically, then one would be able to calculate the pion charge $a_0$ and of course also
higher derivatives. 
 A straightforward calculation gives $a_0 = 1.02 \pm.08 $, 
 where the upper limit of the integration is taken to be $1.7 GeV^2$. The  Ward identity
requires $a_0$ to be exactly unity.  This
 calculation is  to illustrate the fact that the $\rho$ resonance gives the major
contribution and almost saturates the sum rule for the charge; it will not be used in the
following. The higher derivative sum rules are much less sensitive to the high energy
behavior of $ImV(z)$ because of the corresponding weight factor.

It is possible to estimate the high energy contribution to the sum rules by fitting the
asymptotic value of the pion form factor by the expression, $V(s)=-\frac{0.25}{s}
\ln(-s/s_\rho)$ which fits equally well the large energy behavior of the time-like and
space-like pion form factor,  then the integration of the sum rule from $1.7 GeV^2$ to
infinity can be estimated. The r.m.s. radius is then increased by $2\%$ and is completely
negligible for the values of $c$ and
$d$. (The additional contribution to the $a_0$ sum rule is $15\%$).

Notice that the determination of the pion r.m.s. agrees well with its direct experimental
value of $0.439\pm.008 fm^2$ \cite{na7} and with comparable errors. The determination of the
values of
$c$ given by emperical fits to the data using the vector meson dominance models, quoted
as having a large  error of $30\%$, is due to the use of  models  having 
different values for the form factor at the
$\rho$ mass \cite{Gasser3, aleph}. The present 
determination of $c$ as well as of $d$ has an error of only  a few percents.

It is now possible  to compare these experimental results with existing 
calculations.

The first model is obtained using the inverse amplitude method for the pion form
factor at one-loop level. It can also be obtained by using the (0,1)  Pad{\'e} 
approximant method for the one-loop ChPT calculation. The result is well known: by
fitting the time like the pion form factor using the experimental $\rho$ mass, the r.m.s.
radius is 10\% too small. The calculated $\rho$ width is $0.156 GeV$ which is
satisfactory but the maximum value of square of the modulus of the pion form factor is
only 31 which is too low compared with the experimental data value of $40.5\pm0.6$ as given
directly by the $\tau$ decay data.

This result is based on the assumption  of the elastic unitarity which should be correct
for an energy below 1.3 GeV but is certainly incorrect above this energy. One should
phenomenologically correct for this strong assumption. To do this one can simulate the
inelastic effect (and possibly the two-loop singularity) by the presence of the polynomial
ambiguity in the phase representation of the form factor i.e. a zero in the form factor.
Instead of Eq. (\ref{eq:vu}), the time-like pion form factor data can now be fitted by
\cite{Truong4} :
      \begin{equation}
         V(s) = \frac{1+\alpha s/s_\rho} {1 -s/s_{R} - {1\over
96\pi^2f_\pi^2}\{(s-4m_\pi^2)
 H_{\pi\pi}({s}) + {2s/3}\}} \label{eq:vu1}
\end{equation}
where $f_\pi=0.093 GeV$, and $s_{R}$ is related to the $\rho$ mass squared $s_\rho$ by 
requiring that the real part of the denominator of Eq. (\ref{eq:vu1}) vanishes at the $\rho$
mass.

The experimental data can be  fitted with a $\rho$ mass equal to $0.773 GeV$ and
$\alpha=0.14$. This is a two parameter fit to the experimental data and there are three
predictions: the pion r.m.s. radius, the $\rho$ width and the value of the form factor at
the $\rho$ mass (or the $\rho$ leptonic width). These results are in excellent agreement
with the data \cite{aleph,na7}:
 the $\rho$ width, defined as the derivative of the phase shift at the $\rho$ mass, is equal
to $0.156 GeV$, the rms radius is predicted to be $0.45\pm 0.01 fm^2$  and the maximum value
of the modulus squared of the pion form factor at the $\rho$ resonance is 39.2.

The second model, which is more complicated, but is more complete because it is 
based on the two-loop ChPT calculation with unitarity taken into account. It has the
singularity associated with the two loop graphs. By using the same inverse amplitude method
as was done with the one-loop amplitude, but by generalizing this method to two-loop
calculation, Hannah has recently obtained a remarkable fit to the pion form factor in the
time-like and space-like regions. His result is equivalent to  the (0,2)  Pad{\'e} approximant
method as applied to the two-loop ChPT calculation \cite{hannah1}. Unlike Eq. (\ref{eq:vu1})
which is a two parameter fit,  Hannah calculation is  
 a three parameter calculation which gives the correct maximum value of the pion
form factor, the
$\rho$ mass and width and the pion r.m.s. radius.

As can be seen from Figs. 1 and 2, the imaginary and real parts of these two models are
very much in agreement with the data. A small deviation of $ImV(s)$  above $0.9 GeV$ is due
to a small deviation of the phases of
$V(s)$ in these two models from the data of the P-wave $\pi\pi$ phase shifts.  They both
give also correct results for
$c$ and
$d$ as given by the sum rules, Eq.(\ref{eq:rvn}). For $0<s<0.4 GeV^2$ the real parts of
$V(s)$ of these two models  also agree very well with  its Taylor's series expansion, using
its first 4 terms (not shown). 

The results of the imaginary parts of
the one and two-loop ChPT calculations are also shown in Fig. 1; it is seen that  they differ
significantly from the data.  At low energy, ChPT results for the real parts 
 are not  bad  
as can be seen from Fig. 2. This is due to the dominance of the 
real  subtraction constants (uncalculable in ChPT scheme) and cannot be used to support  the
validity of the ChPT as will be discussed below.

At  one-loop level, ChPT cannot be used to calculate the rms radius because the loop
integral for this quantity is divergent. Using the measured r.m.s. as an input, the one
loop ChPT for the pion form factor is given by Eq. (\ref{eq:pertv}). The coefficient
$c^{ChPT}_1$, where the subscript refers to the one-loop level is 
$c^{ChPT}_1=(960\pi^2f_\pi^2m_\pi^2)^{-1}\simeq 0.626 GeV^{-4} $
and  $d^{ChPT}_1=(13440\pi^2f_\pi^2 m_\pi^4)^{-1} \simeq 2.30 GeV^{-6}$ 
The value for $c$ calculated by the one-loop ChPT is in agreement with a previous
determination  \cite{hannah1} and is too small by a factor of 7 compared with the sum rule
value Eq. (\ref{eq:rvn}); 
$d$ is also a factor of 4 too small compared with the sum rule value, Eq. (\ref{eq:rvn}).

At  two-loop level \cite{Gasser3, Colangelo}, ChPT cannot be used to calculate the expression
for $c$ because of the degree of  divergence of the loop integral. One can calculate,
however, the expression for $d$. Although an analytical formula can be given, it is simpler
to give it numerically: $ d^{ChPT}_2= 4.1 GeV^{-6}$ 
which is a factor of 2.5 too small compared with the sum rule value, Eq. (\ref{eq:rvn}).
The disagreement with the sum rule value is now less than that from the one-loop calculation.

 In the following, one follows the standard dispersion
relation analysis of the data, i.e. given the imaginary part of an amplitude, one can
calculate its real part by dispersion relation.
For this purpose, it is important to realize that the equivalent use of the  Taylor's
series is  the dispersion integral (DI)  on the R.H.S.  of Eq.
(\ref{eq:ff1}). Because of the use of the elastic unitarity, this analysis is at best valid
to a maximum of energy of
$0.8-0.9 GeV$. 

  One should compare  the experimental values of the real part of the DI with
those given by models,   because they are the direct results of the calculation schemes,
unmasked by the dominance of the subtraction constants at low energy. The real  parts of the
DI can be  calculated using 
$ImV(s)$ from  experimental data,  models 1, 2 and ChPT results. In Fig. 3, for clarity, only
the two-loop ChPT calculation is plotted. It is seen
that the  two-loop ChPT results, are  too small compared with the corresponding real parts
calculated from the data, not only for small  s  but also for large s  (using
$n=3$ for the DI). 
At very low energy one recovers the results of the Taylor's series as discussed above.  

The  real parts of the DI,  calculated using models 1 and 2, are in
good agreement with the data at low energy, but show a small deviation from the data above
$0.6 GeV$. This result is expected  because models 1 and 2 violate the phase theorem by a
small amount above $0.9 GeV$, as mentioned above.  

It is seen that perturbation theory is inadequate for the vector pion form factor
calculation for an energy below the $\rho$ mass and even at the point  $s=0$  which
are fairly far away from the
$\rho$ mass.  It would be useful to ask how large
the hypothetical 
$\rho$ mass would be in order that perturbation theory could be trusted. It is difficult to
answer this question in general but one can take the large $N_f$ model \cite{will}  as a
guide, where
$N_f$ is the number of flavors. In this model, the explicit solution for the form
factor as well as for the scattering amplitude can be given. The expression for the form
factor is given by Eq. (\ref{eq:vu}). In order that perturbation theory to be valid, it is 
required that the coefficient $c$ calculated by the ChPT to be much larger than that given
by  the large
$N_f$ model. This condition yields:
\begin{equation}
s_R >>\sqrt{960}f_\pi m_\pi \simeq 1.26 GeV^2 \label{eq:cond}
\end{equation}
or roughly  $s_\rho>>1.26 GeV^2$, a condition  cannot be satisfied by the physical
value of the $\rho$ mass.

The situation may not be as bad
for the scalar form factor as can be seen by considering also the large $N_f$ model.
Instead of Eq. (\ref{eq:cond}) we now have a condition on the scalar r.m.s. of the pion,
$1/6<r^2_{scalar}>=1/s_R^{scalar}$:
\begin{equation}
s_R^{scalar}>>\sqrt{1920/19}\pi f_\pi m_\pi \simeq 0.41 GeV^2 
\label{eq:scalar}
\end{equation}
Using the "experimental" value of the scalar r.m.s. radius, $<r^2>=0.6fm^2$ \cite{GL2,
Gasser3}, one has
$s_R^{scalar}
\simeq 0.40 GeV^2$ which is equal to the R.H.S. of Eq. (\ref{eq:scalar}), instead
of being much larger. The situation is better here than the vector case. It might be barely
possible to apply the ChPT for the scalar form factor but the accuracy of the perturbative
approach could be questionable.

In conclusion, because of the inedequacy of the perturbative loop calculations, one could try
two different approaches to this problem. A more radical approach consists in considering
effective lagrangians just as it was invented for, i.e.  low energy theorems; instead of
doing perturbative loop calculations, one could try to analytically continue these low energy
theorems to the time-like region with the technique of dispersion relation and the constraint
of
 unitarity. A more standard approach is to do perturbative calculation but
resumming the perturbative series in order to satisfy  unitarity. In
both approaches, one could use for example the inverse amplitude, N/D and  Pad{\'e}
methods \cite{Truong1, Truong3}.
 This last method  \cite{Samuel}, based on the results of the perturbation
calculation and 
 briefly reviewed here, could be an useful method but could also give a wrong answer in a
number of problems if care was not taken. In general, one does not have a unique
prescription of how to handle the strong interaction problem.

The author would like to thank Dr. Torben Hannah for a detailed explanation of his
calculation of the two-loop vector pion form factor and also for a discussion of the
experimental situation on the pion form factor data. Useful conversations with T. N. Pham are
acknowleledged.

\newpage

{\bf Figure Captions}

Fig.1~: The imaginary part of the vector pion form factor $ImV(s)$, given by Eq.
(\ref{eq:eu}), as a function of the energy in the $GeV$ unit. The experimental data of the
modulus of the form factor and the P-wave $\pi\pi$ phase shifts are taken from the refences
\cite{barkov,aleph,proto,hyams,martin}. The solid curve  is the the experimental results
with experimental errors; the long-dashed curve is the two-loop ChPT calculation, the medium
long-dashed curve is the one-loop ChPT calculation, the short-dashed curve is from the
modified unitarized one-loop ChPT calculation, Eq. (\ref{eq:vu1}), and the dotted curve is
the unitarized two-loop calculation of Hannah \cite{hannah1}.

Fig. 2~: The real parts of the pion form factor as a function of energy. The curves are as in
Fig. 1.

Fig. 3~: The real parts of the dispersion integral ReDI as a function of energy (using $n=3$)
in Eq. (\ref{eq:ff1}). The curves are as in Fig. 1; the ChPT one-loop result is not shown.

\newpage
\begin{figure}
\epsfbox{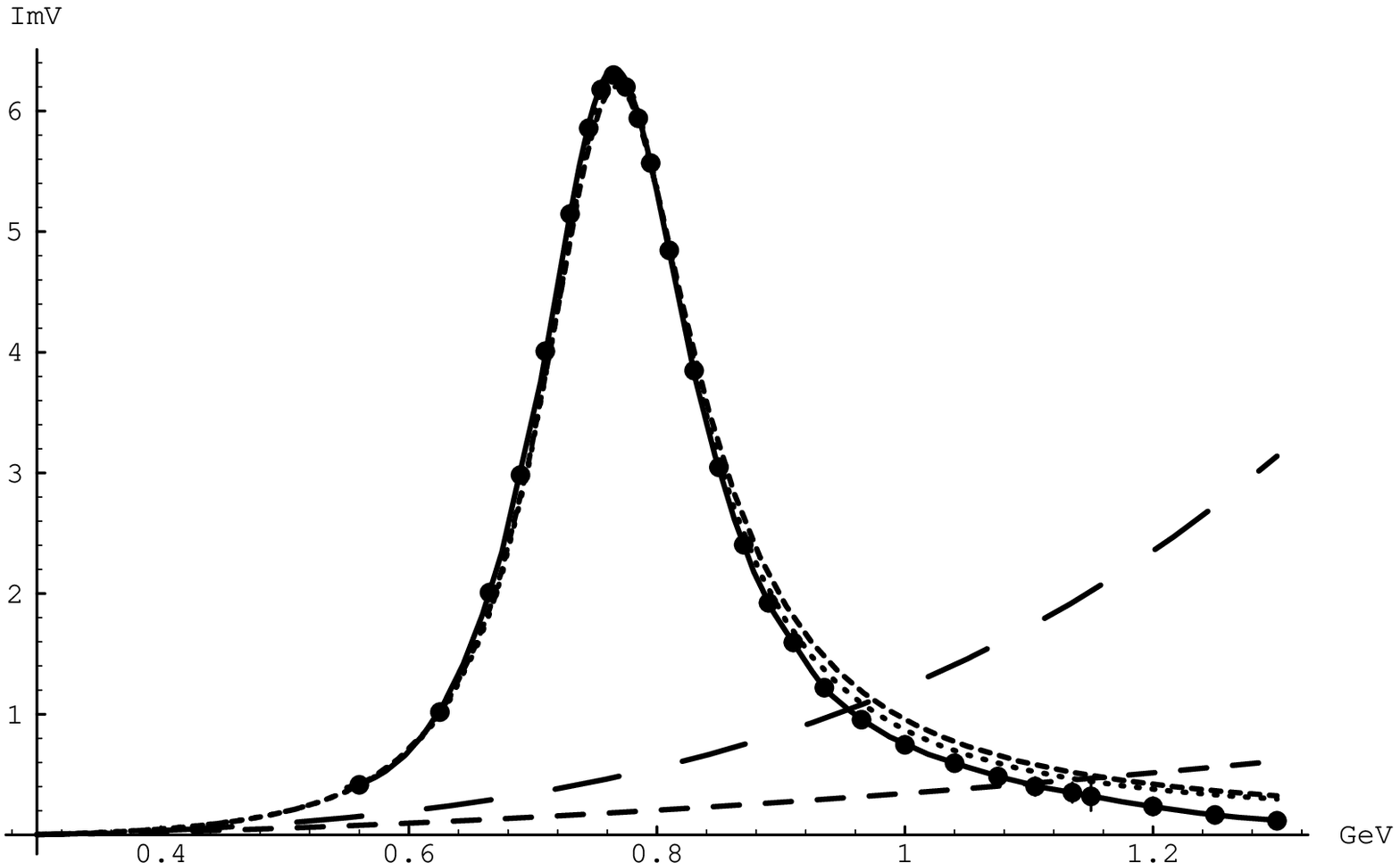}
\caption{}
\label{Fig.1}
\end{figure}
\newpage
\begin{figure}
\epsfbox{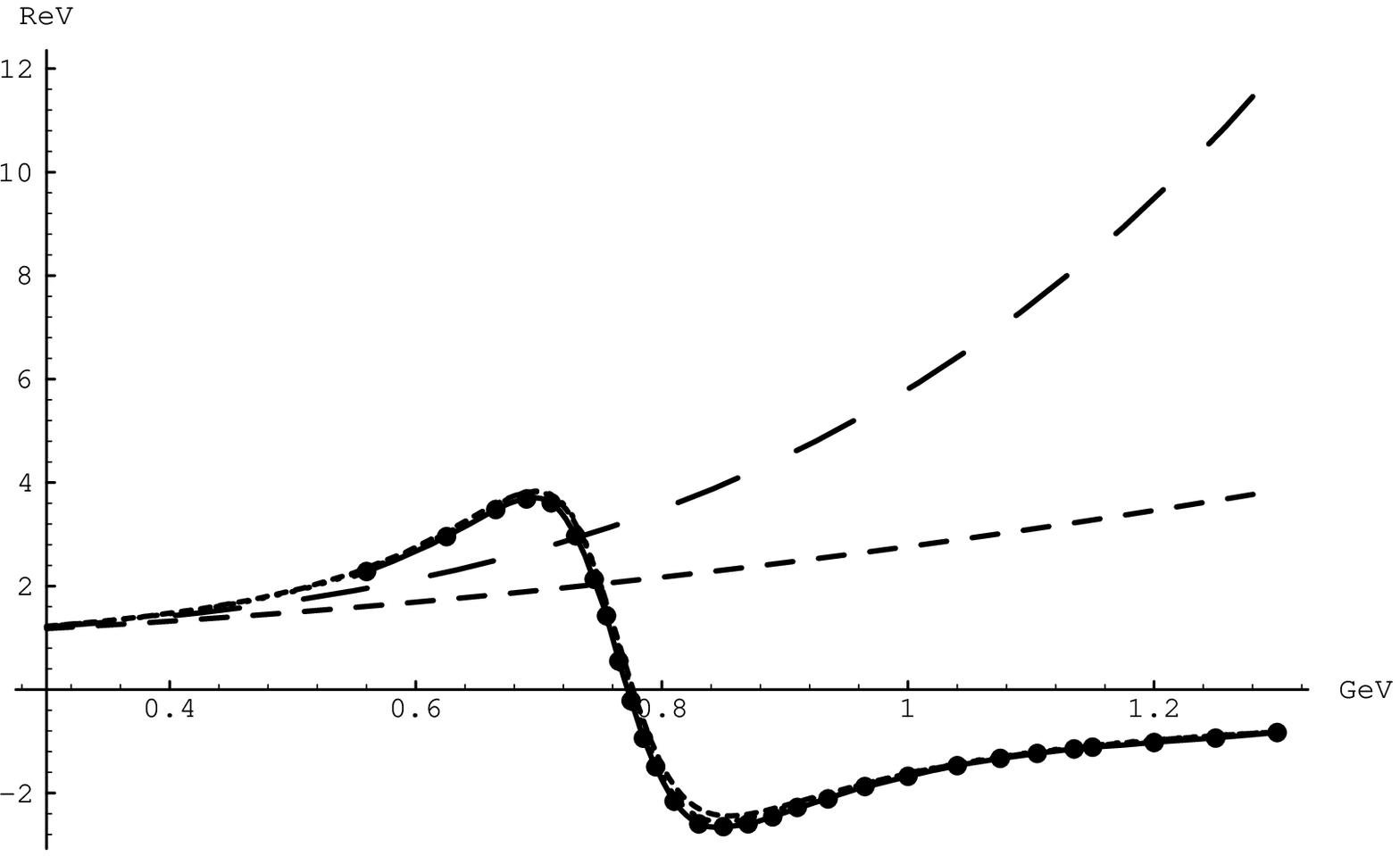}
\caption{}
\label{Fig.2}
\end{figure}
\newpage
\begin{figure}
\epsfbox{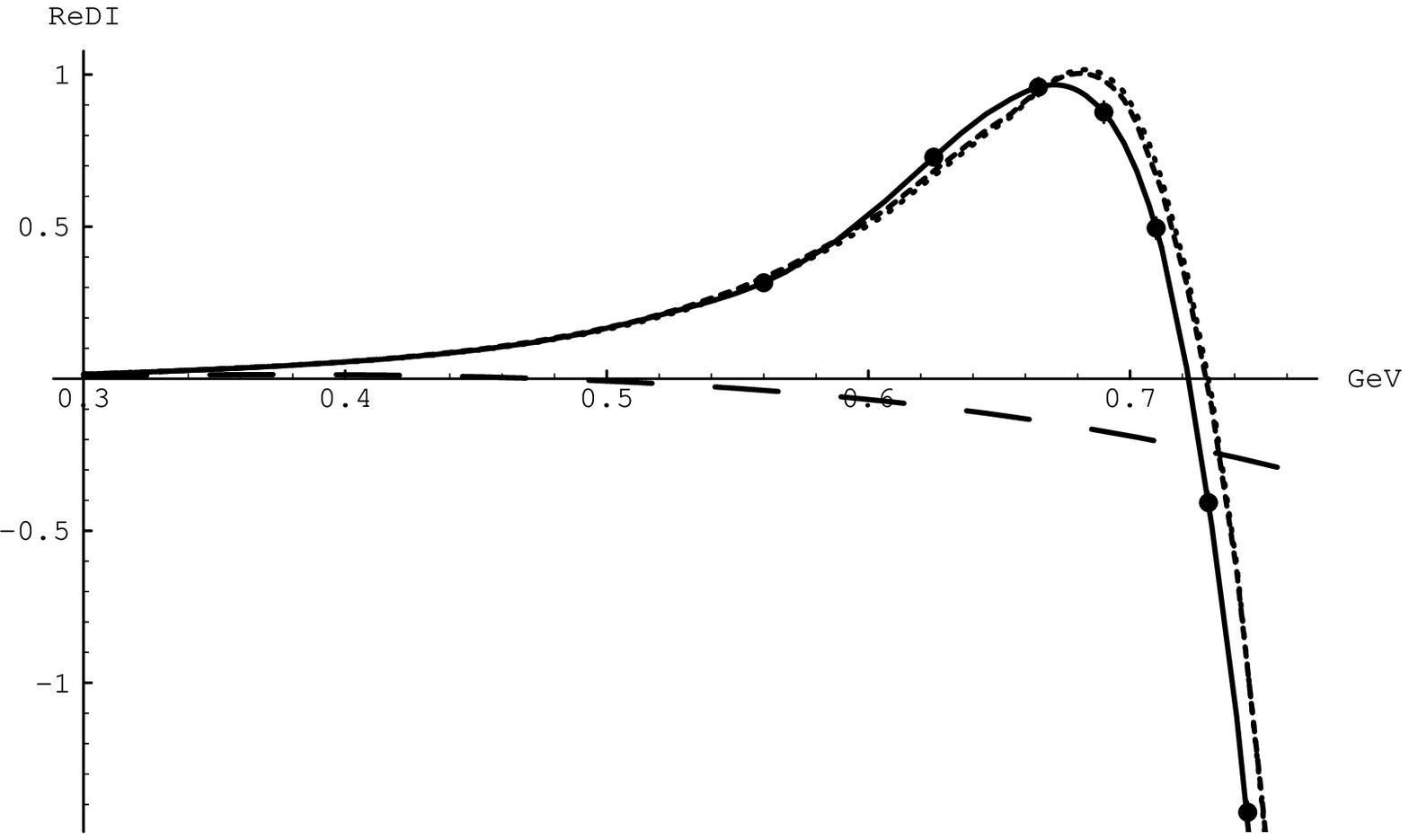}
\caption{}
\label{Fig.3}
\end{figure}


\newpage
\begin{thebibliography}{99}

\bibitem{holstein} J. F. Donoghue, E. Golowich and B. R. Holstein, \emph {Dynamics of the
Standard Model} (Cambridge Univ. Press, Cambridge, 1992).
\bibitem{Weinberg} S. Weinberg, Physica 96 A (1979) 327.
\bibitem{GL1} J. Gasser and H. Leutwyler, Ann. Phys. (N.Y.) {\bf 158}
  142 (1984).
\bibitem{GL2} J. Gasser and H. Leutwyler, Nucl. Phys. {\bf B250} 465 (1985) .\\
ibid. {\bf B250} 539 (1985).
\bibitem{Gasser3} J. Gasser and Ulf-G Mei$\beta$ner, Nucl. Phys. {\bf B357} 90 (1991).
\bibitem{Colangelo} G. Colangelo, M. Finkemeier and R. Urech, Phys. Rev. {\bf D54}, 4403
(1996).
\bibitem{Truong1} T. N. Truong, Phys. Rev. Lett. {\bf 61}, 2526 (1988).
\bibitem{Truong3} T. N. Truong, Phys. Rev. Lett. {\bf 67}, 2260 (1991).
\bibitem{watson} K. M. Watson, Phys. Rev. {\bf 95}, 228 (1954).
\bibitem{oller}F. Guerrero and J. A. Oller,  Report No hep-ph/9805334.
\bibitem{barkov} L. M. Barkov \emph{et al.} Nucl. Phys. {\bf B256}, 365 (1985).
\bibitem{aleph} ALEPH Collaboration, R. Barate \emph{et al.}, Z. Phys. C {\bf 76}, 15
(1997). 
\bibitem{proto} S. D. Protopopescu \emph {et al.}, Phys. Rev. D {\bf 7}, 1279 (1973).
\bibitem{hyams} B. Hyams \emph{et al.} Nucl. Phys. {\bf B64}, 134 (1973).
\bibitem{martin} P. Eastabrooks and A. D. Martin, Nucl. Phys. {\bf B79}, 301 (1974).
\bibitem{na7} NA7 Collaboration, S. R. Amendolia \emph{et al.}, Nucl. Phys. {\bf B277}, 168
(1986).
\bibitem{Truong4} Le viet Dung and Tran N. Truong, Report No hep-ph/9607378.
\bibitem{hannah1} T. Hannah, Phys. Rev. {\bf D55}, 5613 (1997). Ph. D thesis, Aarhus
University (1998).
\bibitem{Truong2} A. Dobado, M.J. Herrero and T. N. Truong, Phys. Lett.
 {\bf B 235} 129, 134 (1990).
\bibitem{will} S. Willenbrock, Phys. Rev. {\bf D43},1710 (1991) and references cited therein.
\bibitem{dobado} 
A. Dobado and J. R. Pelaez, Phys. Rev. {\bf D56} 3057 (1997).
\bibitem{Samuel} For other applications of the Pad{\'e} method see: M.A. Samuel, J. Ellis and
M. Karliner, Phys. Rev. Lett. {\bf 74}
 (1995) 4380;\\ S. J. Brodsky, J. Ellis, E. Gardi, M. Karliner, M. A. Samuel, Phys. Rev.
{\bf D56} 6980 (1997).


\end{thebibliography}
\end{document}